\documentclass{PoS}
\usepackage{cite}
\title{Directed flow in Au+Au collisions from the RHIC Beam Energy Scan at
  the STAR experiment}

\ShortTitle{Directed flow in Au+Au collisions from the RHIC Beam Energy Scan at STAR}

\author{\speaker{Subhash Singha}\\
  (for the STAR Collaboration)\\
        Kent State University, Ohio 44242, USA\\
        E-mail: \email{subhash@rcf.rhic.bnl.gov}}

\abstract{
We report results of $v_1(y)$ and $dv_1/dy$ near mid-rapidity
for $\pi^{\pm}$, $K^{\pm}$, $K_s^0$, $p$, $\overline{p}$, $\Lambda$,
$\overline{\Lambda}$ and $\phi$ from Beam Energy Scan Au+Au collisions at 
$\sqrt{s_{NN}} = $~7.7 -- 200~GeV using the STAR detector at RHIC. The $dv_{1}/dy$ of
$\pi^{\pm}$, $K^{\pm}$ and $K_s^0$ mesons
remains negative over all beam energies. The $dv_1/dy$ of
$p$ and $\Lambda$ baryons shows a sign change around 10 -- 15~GeV, while
net baryons (net~p and net~$\Lambda$) indicate a double sign
change. The $dv_1/dy$ of $\overline{p}$, $\overline{\Lambda}$
and $\phi$ show a similar trend for $\sqrt{s_{NN}}>$ 14.5~GeV.  
For the first time, $v_{1}$ measurements are used to test a quark
coalescence hypothesis. Many measurements are found to be consistent with
the particles being formed via coalescence of constituent
quarks. The observed deviations from that consistency 
offer a new approach for probing the collision process at the quark level.} 

\FullConference{Critical Point and Onset of Deconfinement - CPOD2017\\
		7-11 August, 2017\\
		The Wang Center, Stony Brook University, Stony Brook, NY}

\begin{document}

\section{Introduction}

A goal of research at the Relativistic Heavy Ion Collider (RHIC) is to
explore deconfined quark-gluon matter~\cite{STARwhitepaper}. Directed flow ($v_{1}$) is
one of the observables that is sensitive to the dynamics of the system
in early times during the collision process. Both hydrodynamical and transport
model calculations indicate that the $v_{1}$ slope at mid-rapidity,
especially for baryons, is sensitive to the Equation of State (EoS)
of the system \cite{Rischke, Stoecker}. Based on hydrodynamics, 
a minimum in $v_{1}$-slope is proposed as a signature
of a first-order phase transition between hadronic matter and
quark-gluon plasma (QGP)~\cite{Rischke, Stoecker}. STAR has taken data over a wide range of beam
energies ($\sqrt{s_{NN}} = $7.7--200~GeV) to explore the QCD phase diagram. One of the
goals of this Beam Energy Scan (BES) is to find the softest point
of the QCD Equation of State~\cite{BES-I}. The $v_1$-slope of protons from 
phase-I of BES~\cite{STAR-BESv1} changes sign near $\sqrt{s_{NN}} \sim 10$~GeV with a 
minimum around $\sqrt{s_{NN}} = $10--20~GeV. Also the net-proton $v_1$-slope shows a double 
sign change with a more pronounced minimum in the same energy range. 
Such a behavior shows a qualitative resemblance to a
3-fluid hydrodynamic model calculation with a 1st-order phase transition~\cite{Stoecker}. However, current 
state-of-the-art models are not able to reproduce the basic trend of energy dependence of proton
$v_1$-slope reported by STAR~\cite{hybrid, phsd, phsd2, 3FD, JAM, JAM2}. Moreover, the $v_{1}$ results from 
models with nominally similar prescriptions for the equation of state differ by an order of magnitude~\cite{v1review}. 
More theoretical progress is necessary for a definitive interpretation of the data.

Recently we have measured $v_1$ for $\Lambda$, $\bar{\Lambda}$, $K^{\pm}$, $K_{S}^{0}$
and $\phi$ in Beam Energy Scan Au+Au collisions at $\sqrt{s_{NN}} =
$~7.7--200~GeV using the STAR detector~\cite{v1ncq}. $\Lambda$ hyperon offers the
opportunity to study a second baryon species, and can compliment the
proton data. Charged kaons and $K_{S}^{0}$ have been observed to depend on the kaon-nucleon potential at the energy of 
AGS experiments \cite{kaon-potn}. One can test whether any such effect can be observed
in the low energy domain of BES. Furthermore, the $\phi$ meson is interesting because its
mass is close to the mass of baryons, but it is a vector meson.  So
$\phi$ can serve as a probe to study whether the $v_{1}$ depends on
flavor (baryon or meson) or mass of a particle species.
The $\phi$ is less affected by hadronic interactions than other species.
Thus it can be used as a clean probe to study
the contribution of the partonic phase to $v_{1}$. The $v_1$ measurements with ten
different hadron species, having different constituent quarks, will
help to disentangle the role of produced and transported quarks in heavy-ion collisions.

\begin{figure}
\begin{center}
\includegraphics[scale=0.8]{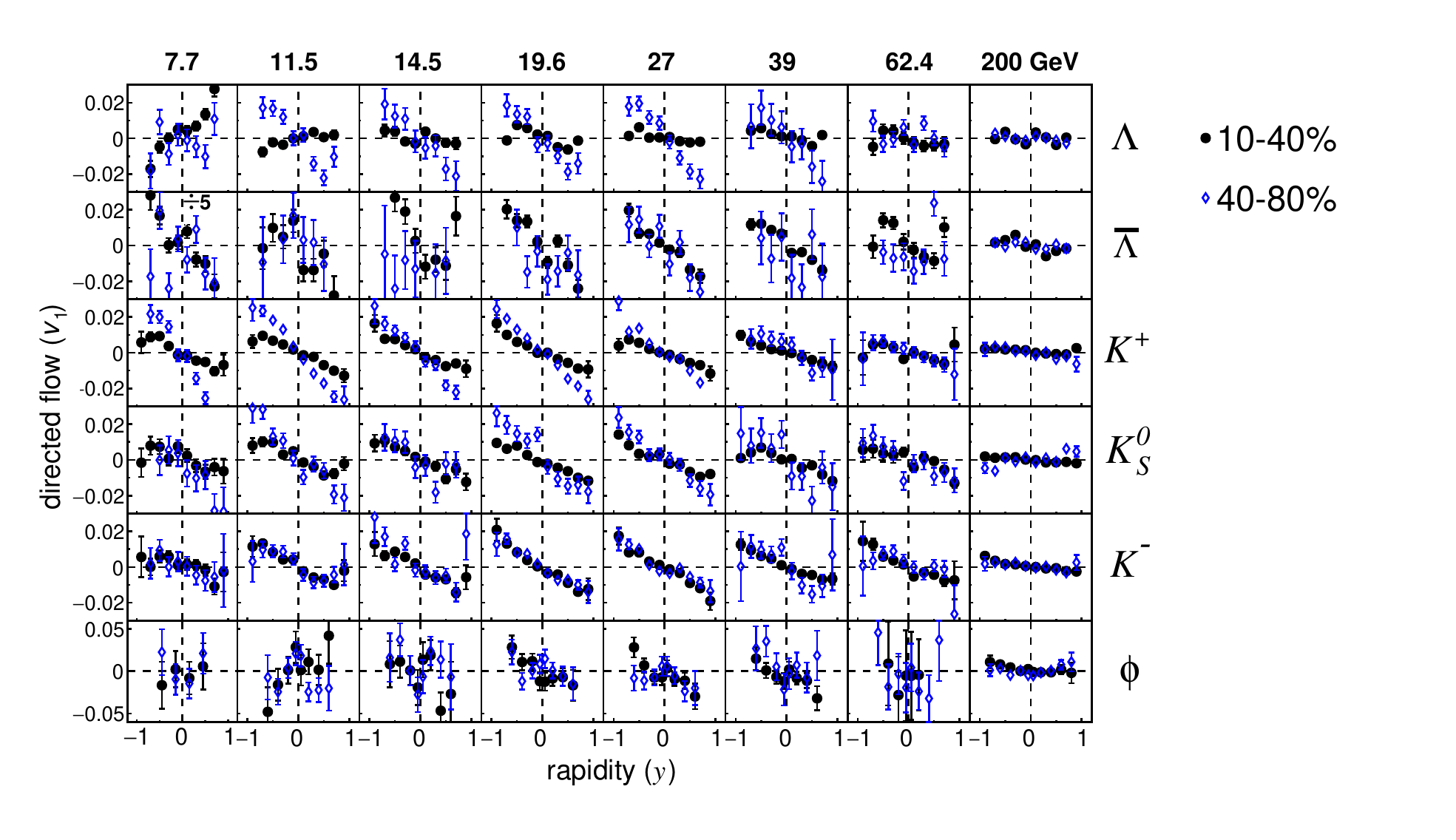}
\caption{(Color online) Rapidity dependence of directed flow ($v_{1}$)
for $\Lambda$, $\overline{\Lambda}$, $K^{+}$, $K_s^0$, $K^{-}$ and $\phi$
in 10-40\% and 40-80\% Au+Au collisions at $\sqrt{s_{NN}} = $  7.7,
11.5, 14.5, 19.6, 27, 39, 62.4 and 200~GeV.}
\label{fig-v1vsy}
\end{center}
\end{figure} 
\section{Analysis Details}
The analysis was done using Au+Au collision data taken by
the STAR detector during the years 2010, 2011, and 2014. 
The STAR detector offers uniform acceptance, full azimuthal coverage and
excellent particle identification~\cite{starnim}. The Time Projection Chamber (TPC)~\cite{startpc} 
is the main detector which performs charged particle tracking near
mid-rapidity. The collision centrality is estimated using the charged particle multiplicity in the 
rapidity region $|\eta| < 0.5$, measured with the TPC. We utilize both the Time Projection Chamber
(TPC) and Time-of-Flight (ToF) ~\cite{TOF} detectors for particles identification. 
The first-order reaction plane ($\Psi_1$) is
estimated using the two Beam Beam Counters (BBC)~\cite{BBC}, covering 3.3 < $|\eta|$
< 5.2,  for $\sqrt{s_{NN}} = $~ 7.7--39~GeV. The event plane
resolution from the BBCs deteriorates at $\sqrt{s_{NN}}$  > 39~GeV. So we
utilize the Zero Degree Calorimeters (ZDC)~\cite{GangThesis}, which cover $|\eta| >
6.3$, to estimate $\Psi_1$ for $\sqrt{s_{NN}} = $~62.4 and
200~GeV. The large $\eta$-gap of these forward detectors (BBC and ZDC) relative to the
TPC reduces non-flow contribution in our $v_{1}$ measurements. 

\section{Results and discussions} 
\begin{figure}
\begin{center}
\includegraphics[scale=0.5]{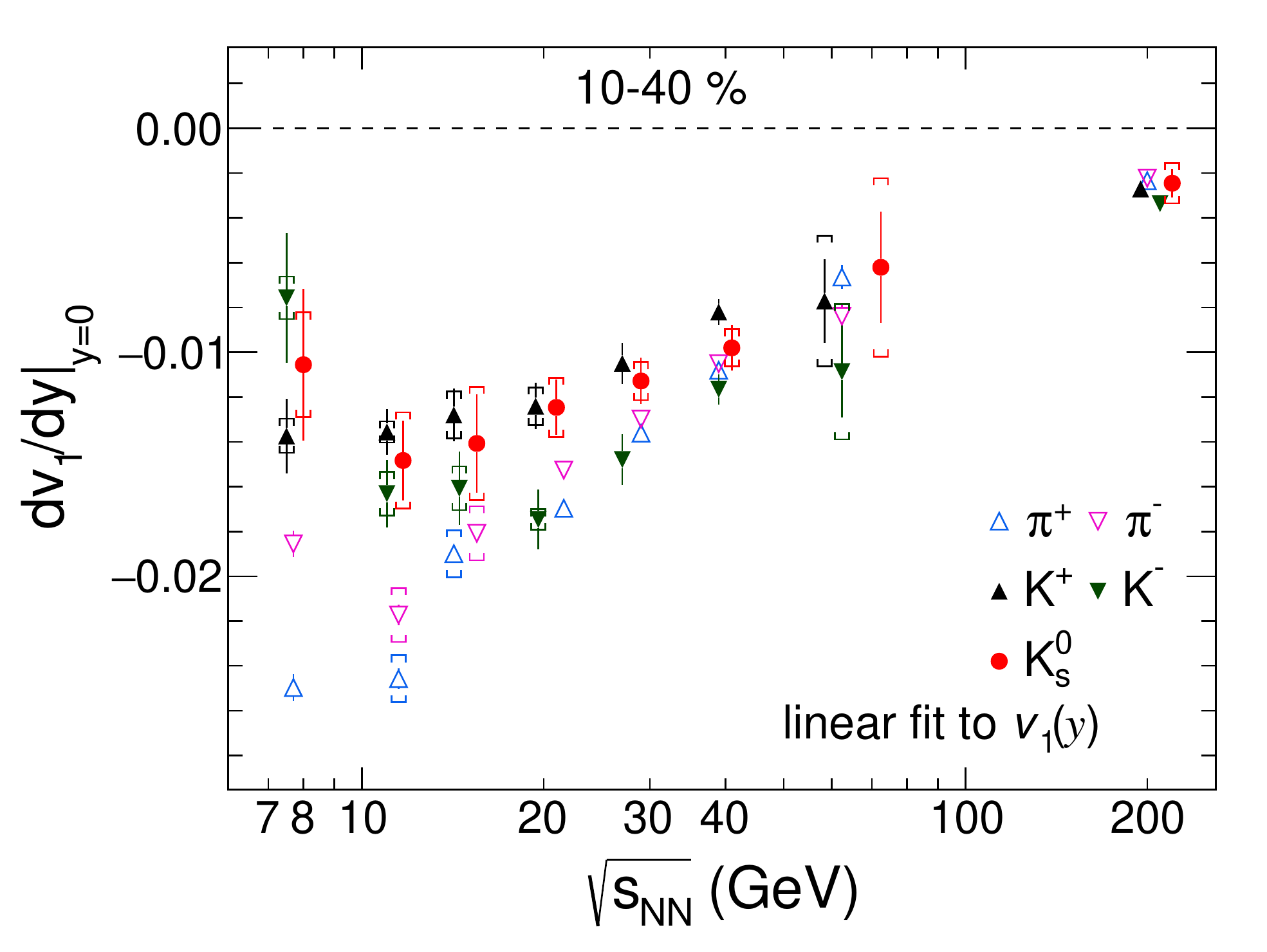}
\includegraphics[scale=0.5]{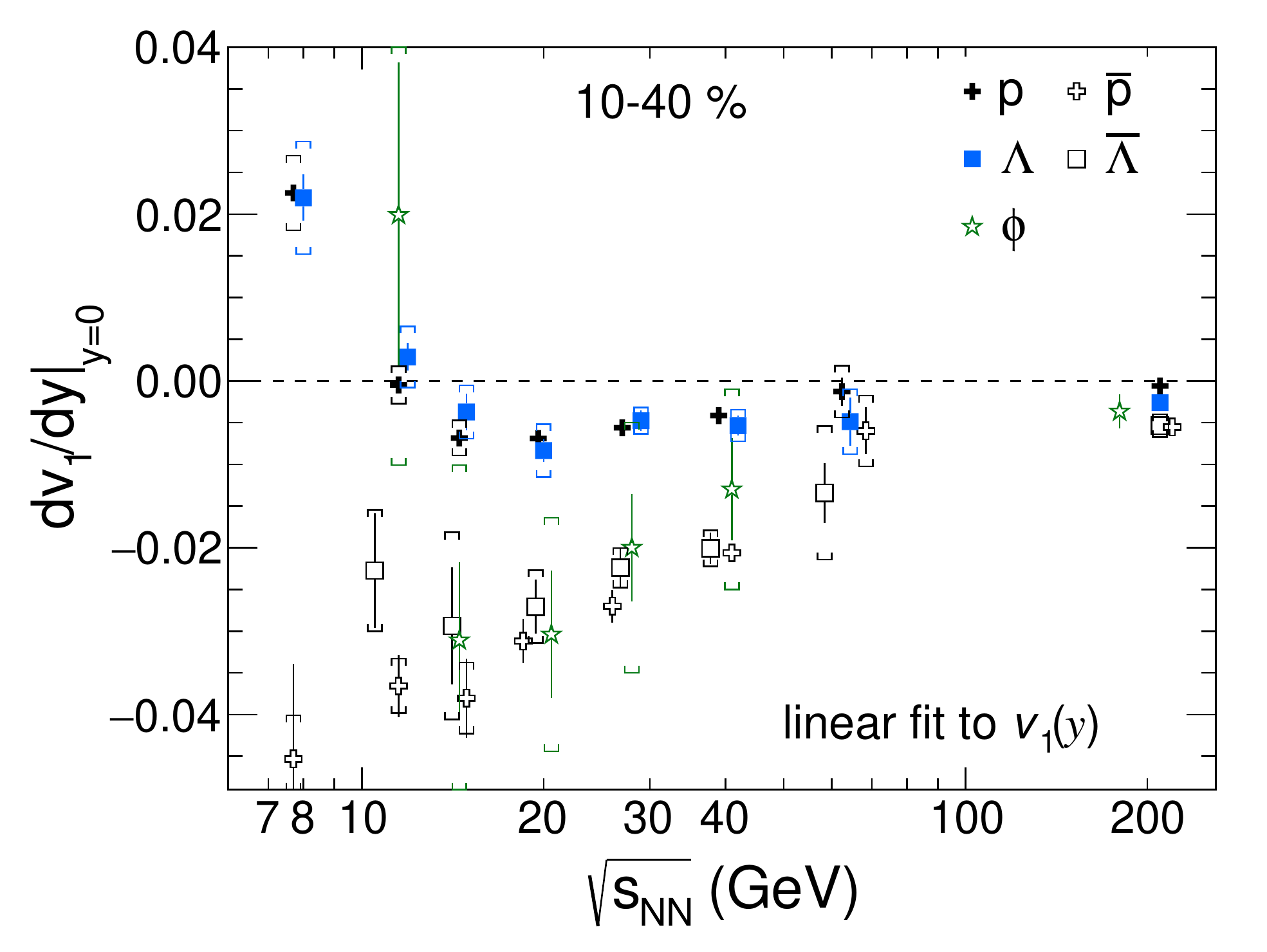}
\caption{(Color online) Top panel: Beam energy dependence of
  $dv_{1}/dy$ for $\pi^{\pm}$, $K^{\pm}$ and $K_s^0$ in 10-40\% Au+Au
  collisions. Bottom panel: Beam energy dependence of
  $dv_{1}/dy$ for $\Lambda$, $\overline{\Lambda}$, $p$, $\bar{p}$ and $\phi$ in 10-40\% Au+Au
  collisions.}
\label{fig-dv1dy}
\end{center}
\end{figure}

Figure~\ref{fig-v1vsy} reports the rapidity dependence of $v_{1}$ for
$\Lambda$, $\overline{\Lambda}$, $K^{\pm}$, $K_{S}^{0}$ and $\phi$ in
$10-40\%$ and $40-80\%$ central Au+Au collisions at
$\sqrt{s_{NN}} = $~7.7, 11.5, 14.5, 19.6, 27, 39, 62.4 and 200~GeV. The
data points for $\overline{\Lambda}$ at 7.7~GeV are divided by five in
order to present the data in the same vertical scale. It is a common practice to present
the strength of $v_{1}(y)$ by its slope at mid-rapidity. In Ref.~\cite{STAR-BESv1}, the $v_1$ slope parameter
($dv_{1}/dy$) at mid-rapidity was extracted by fitting the data to a cubic
function ($F_{1}\,y + F_{3}\,y^{3}$). In this analysis, poor statistics for $\overline{\Lambda}$
and $\phi$ do not allow a stable cubic fit as done in Ref.~\cite{STAR-BESv1}. We extract $dv_{1}/dy$
by using a linear function ($F_{1}\,y$) over the rapidity region $|y|< 0.8$, 
except for $\phi$ where the fit is performed for $|y|< 0.6$.
The energy dependence of the $v_{1}$ slope for $\pi^{\pm}$,
$K^{\pm}$ and $K_{S}^{0}$ mesons are shown in the top panel in Fig~\ref{fig-dv1dy}. It is observed 
that all the mesons have negative $v_{1}$ slope. The magnitude of the slope increases
with decreasing beam energy. The $K^+$
slope lies above $K^-$ at and above $\sqrt{s_{NN}} = $~11.5~GeV while a
reversal of their positions is observed at $\sqrt{s_{NN}} = 7.7$~GeV. The 
$K_{S}^{0}$ is consistent with the mean of $K^{+}$ and $K^{-}$ at all studied 
beam energies. Within the present uncertainties, no
distinct mass ordering is observed among the mesons. 

The energy dependence of $dv_{1}/dy$
for the baryons and for $\phi$ are presented in the bottom panel of Fig.~\ref{fig-dv1dy}. 
We observe that $p$ and $\Lambda$ show a sign change near
$\sqrt{s_{NN}} = $11.5~GeV. 
The $dv_{1}/dy$ of $\bar{p}$ and $\overline{\Lambda}$ remain negative at all beam
energies and their magnitude increases with decreasing beam
energy. The $\phi$ meson $dv_{1}/dy$ follows the trend of
anti-baryons for $\sqrt{s_{NN}} > $ 14.5~GeV, while its slope turns
towards zero at lower energies. Current statistical and systematic
uncertainties for $\phi$ $v_{1}$ at 7.7 and 11.5~GeV are too large to draw any
conclusions. 

\begin{figure}
\begin{center}
\includegraphics[scale=0.5]{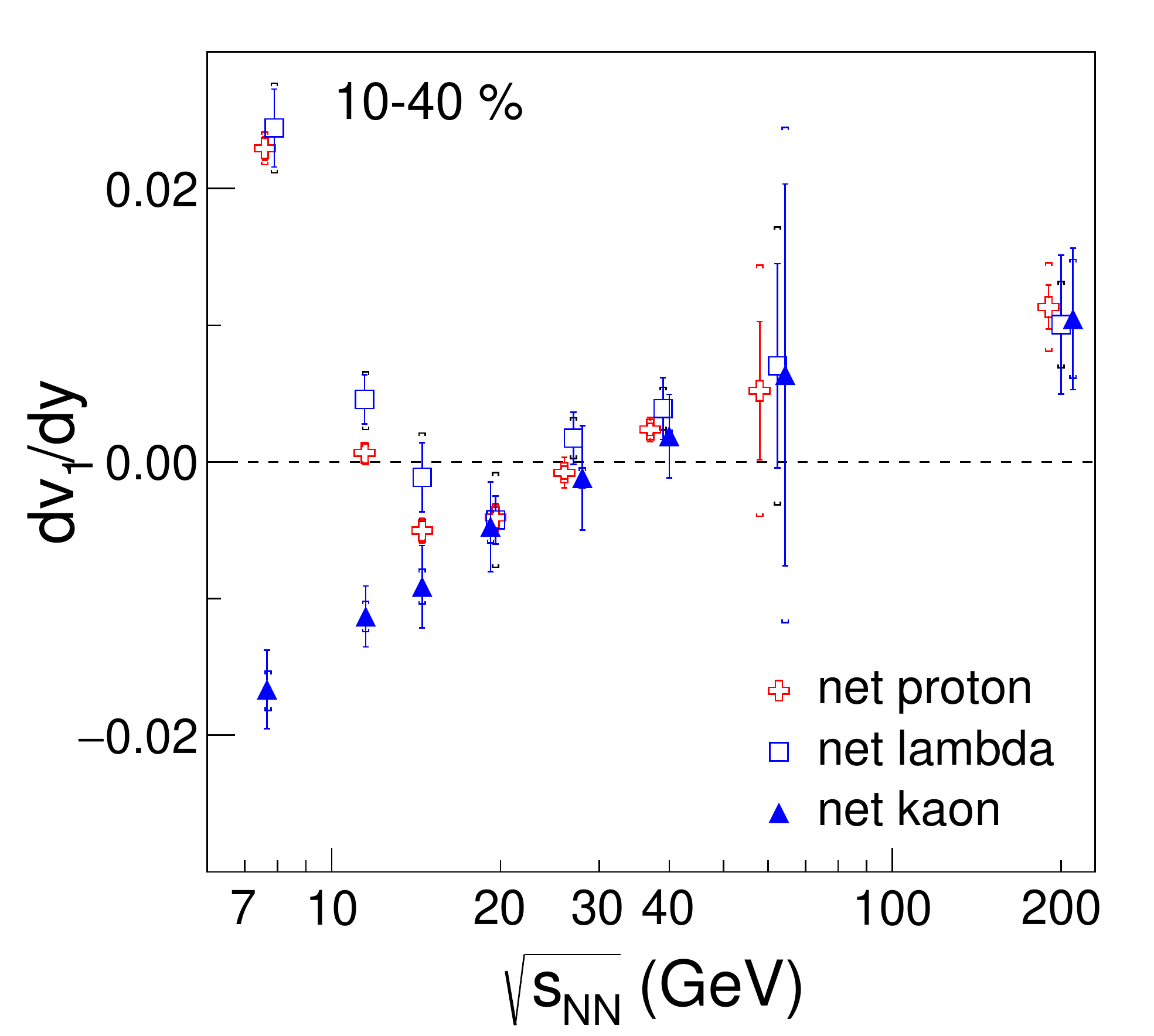}
\caption{(Color online) Beam energy dependence of net-particle
  $dv_{1}/dy$ in 10-40\% Au+Au collisions.}
\label{eg-netpv1}
\end{center}
\end{figure}

The $v_1$ can have contributions from both produced and transported quarks. 
To disentangle the contribution of the two sources, we define
a net particle $v_{1}$ as \\
\begin{eqnarray}
\label{formula1}
F_{\Lambda} = r_{1}(y) \,F_{\,\overline{\Lambda}} + [1 - r_{1}(y)]\,
  F_{{\rm  net\,} \Lambda} ,
\end{eqnarray} 
\begin{eqnarray}
\label{formula2}
F_{p} = r_{2}(y) \,F_{\,\overline{p}} + [1 - r_{2}(y)]\, F_{{\rm
  net\,}p} ,
\end{eqnarray} 
\begin{eqnarray}
\label{formula3}
F_{K^{+}} = r_{3}(y) \,F_{K^{-}} + [1 - r_{3}(y)]\, F_{{\rm net\,}K} ,
\end{eqnarray} 
where $F$ denotes the $dv_{1}/dy$ for each species and $ r_{i}(y)$ denotes
the rapidity dependent ratios of corresponding anti-particle to
particle yields. Figure~\ref{eg-netpv1} presents the energy
dependence of directed flow slope for net $\Lambda$, net $p$ and net $K$. It is observed that
net baryons (net $p$ and net $\Lambda$) agree within uncertainties at all energies.  
The net-$K$ $dv_{1}/dy$ is observed to follow net $p$ and net $\Lambda$ for $\sqrt{s_{NN}}>$ 
14.5~GeV, while it diverges from the trend below 14.5~GeV. 

\begin{figure}
\begin{center}
\includegraphics[scale=0.5]{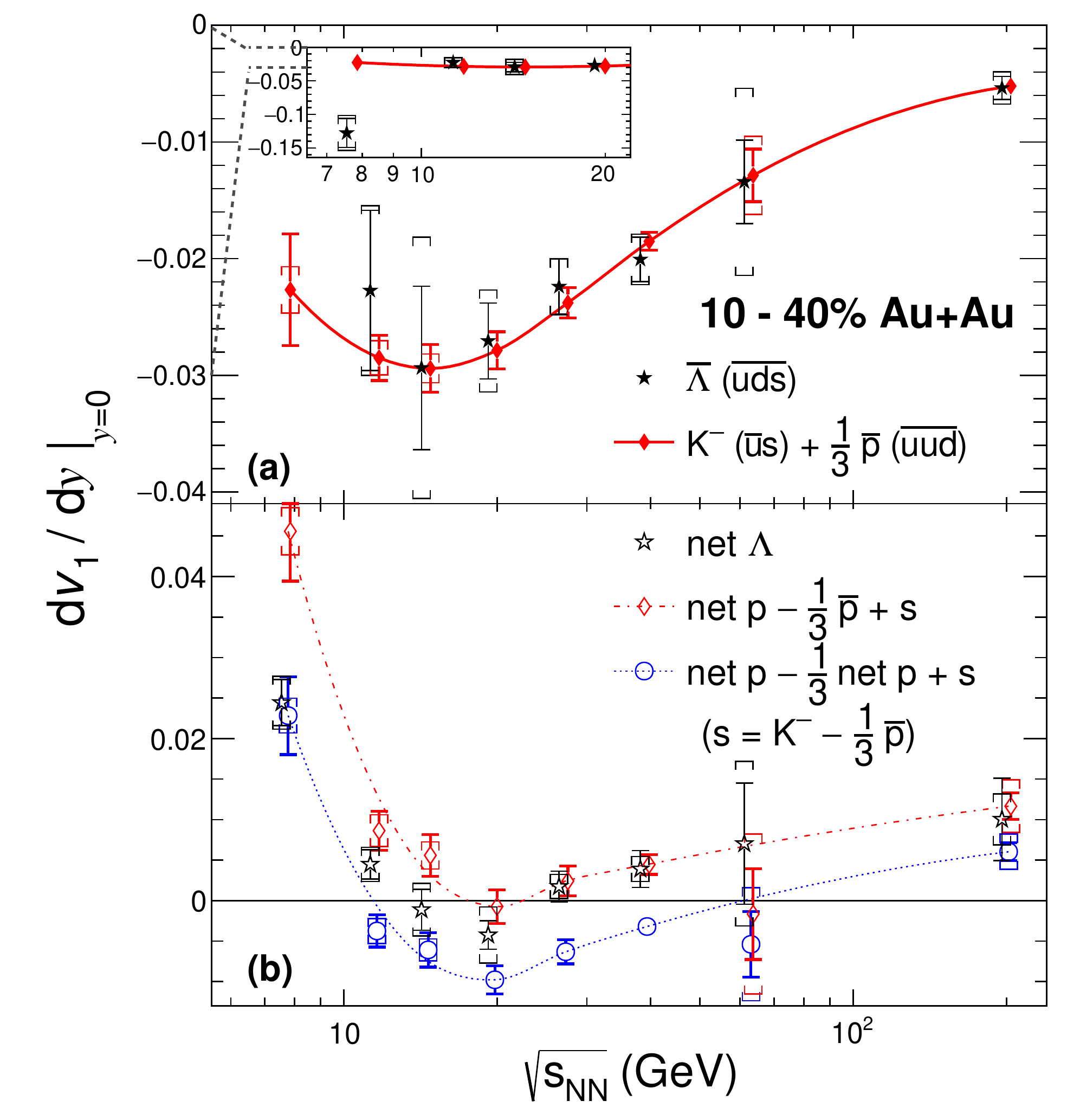}
\caption{(Color online) Directed flow slope ($dv_1/dy|_{y=0}$) versus
  $\sqrt{s_{NN}}$ for intermediate-centrality (10-40\%) Au+Au
  collisions. Panel (a) compares the observed $\bar{\Lambda}$ slope
  with the prediction of the coalescence sum rule for produced
  quarks. The inset shows the same comparison where the vertical scale
  is zoomed-out; this allows the observed flow for the lowest energy
  ($\sqrt{s_{NN}} = 7.7$~GeV) to be seen. Panel (b) presents two
  further sum-rule tests, based on comparisons with net-$\Lambda$ measurements. The expression $K^-(\bar{u}s) - {1 \over 3} \bar{p}\,(\overline{uud})$ represents the $s$ quark flow; there is no corresponding clear-cut expression for transported $u$ and $d$ quarks.}
\label{fig-v1scaling}
\end{center}
\end{figure}

Number of Constituent Quark (NCQ) scaling has been observed in
higher flow harmonics ($v_{2}$ and $v_{3}$) at RHIC and LHC
energies~\cite{ncq1, ncq2, ncq-lhc}. Such scaling is interpreted as evidence of
quark degrees of freedom in the early stages of heavy-ion collisions. 
The $v_{1}$ measurements for ten different particle species allow us to extend tests of a quark
coalescence hypothesis using both produced and transported quarks. In a naive 
quark coalescence picture, we can assume that quarks acquire $v_{1}$ from the medium 
and then those quarks statistically coalesce to form hadrons.  The 
coalescence sum rule hypothesis can be written as
$(v_{n})_{\rm hadron} = \sum  (v_{n})_{\rm constituent \, quarks}$.
In panel~(a) of Fig.~\ref{fig-v1scaling}, we present a test of the quark
coalescence sum rule hypothesis utilizing particle species where all the
constituent quarks are produced in collisions. The solid black marker presents $dv_{1}/dy$ for
$\overline{\Lambda}$. This result is compared with a sum rule
hypothesis where we add one third of $\overline{p}$ $dv_{1}/dy$ to
$K^{-}$ $dv_{1}/dy$.  The factor $\frac{1}{3}$ arises from the assumption
that $\bar{u}$ and $\bar{d}$ have the same $v_{1}$. We also assume
that the $s$ and $\bar{s}$ have the same $v_{1}$. We observe that
this sum rule hypothesis holds for $\sqrt{s_{NN}} = $11.5--200~GeV,
while it deviates at 7.7~GeV. The deviation at the lowest
energy indicates that the above-mentioned assumptions do not hold
at $\sqrt{s_{NN}} = $7.7~GeV.

Panel~(b) in Fig.~\ref{fig-v1scaling} presents tests of the coalescence sum rule hypothesis 
using $u$ and $d$ quarks, which could be transported or produced.  
We utilize net-particle $v_{1}$ to test the coalescence hypothesis in this less straightforward situation.  
The fraction of transported quarks among the
constituent quarks of net particles is larger than in particles roughly in proportion to 
$N_{\rm particle}/N_{\rm net \, particle}$. Here we assume that all the transported quarks are
contained in net particles. In panel~(b) of Fig.~\ref{fig-v1scaling}, open black symbols 
present net-$\Lambda$ $dv_{1}/dy$. The contribution of transported
quarks will increase in net $\Lambda$ at low beam energies, where the $u$
and $d$ quarks are more likely to be transported from the colliding
nuclei. In the high energy limit, the constituent quarks in net $\Lambda$ are more likely to be
produced in the collision. The net-$\Lambda$ result is compared with two different
coalescence sum rule calculations. In the first calculation, shown by open red markers, 
we replace one of the $u$ quarks in net proton with a $\bar{u}$ quark, and also add the 
$dv_1/dy$ contribution from an $s$ quark. The $s$ quark contribution is obtained by subtracting 
one third of $\overline{p}$ $dv_1/dy$ from $K^{-}$ $dv_{1}/dy$. The
first sum rule calculation is consistent with net $\Lambda$ within errors
for $\sqrt{s_{NN}} >$ 19.6~GeV.
As the contribution of transported quarks increases in net $\Lambda$ at lower energies, the
first sum rule calculation deviates as we scan further below
19.6~GeV. In the second sum rule test (shown by blue
open markers), we add one $s$ quark $dv_1/dy$ to two-thirds of net-$p$ 
$dv_{1}/dy$. Here we assume that the constituent quarks in net $p$ are
dominated by transported quarks. We find that this
second sum rule calculation agrees with net $\Lambda$ at the
lowest beam energy and it deviates with increasing beam energy.

\section{Conclusions and outlook}

In summary, we reported the measurement of directed flow ($v_1$) near midrapidity 
for $\pi^{\pm}$, $K^{\pm}$, $K_s^0$, $p$, $\bar{p}$, $\Lambda$,
$\bar{\Lambda}$ and $\phi$ spanning eight beam energies over the range 
$\sqrt{s_{NN}} = $7.7 to 200~GeV. We observe that the proton and
$\Lambda$ $dv_1/dy$ agree within errors and they change sign near
$\sqrt{s_{NN}} = $11.5~GeV. The slopes for $\pi^\pm$, $K^\pm$,
$K^0_s$, $\bar{\Lambda}$ and $\bar{p}$ are negative at all available beam energies. The
$K^0_s$ is consistent with the mean of $K^+$ and $K^-$ at all 
beam energies. The $\phi$ $dv_1/dy$ is negative and follows the same
trend as $\bar{p}$ and $\bar{\Lambda}$ for $\sqrt{s_{NN}}>14.5$~GeV.
Net-particle $dv_1/dy$ for $p$, $\Lambda$ and
$K$ agree with each other at and above $\sqrt{s_{NN}} = 14.5$~GeV, but
net $K$ diverges at 11.5 and 7.7~GeV. 
A quark coalescence sum rule hypothesis is tested using $v_1$ measurements. 
Produced quarks follow coalescence sum-rule behavior at 11.5--200~GeV, but strongly 
deviate at 7.7~GeV. This offers a new approach to
probe the heavy-ion collision process at the quark level. 

The STAR collaboration will upgrade detectors in phase-II of the RHIC Beam
Energy Scan~\cite{BES-II}. The upgrade of the inner TPC will not only improve particle
identification but will also enhance forward rapidity
coverage. A new endcap Time-of-Flight detector (eTOF) will further extend particle
identification. Furthermore, future measurements
will greatly benefit from a new Event Plane Detector (EPD), which will improve 
the first-order event plane resolution by a factor of two. The EPD will also provide an
independent collision centrality determination. These detector
upgrades and increased statistics in BES-II will significantly
improve the precision and quality of many measurements.

Recently, a model calculation~\cite{hfv1} predicted that the early transient 
magnetic field can induce a larger $v_{1}$ for heavy quarks than
for light quarks. The model calculation also suggests an opposite sign
of $dv_{1}/dy$ for charm ($c$)
and anti-charm ($\bar{c}$) quarks. This offers hope that heavy quarks will offer 
a new approach to study the early electromagnetic field. STAR has collected
over 2 billion events with the Heavy Flavor Tracker (HFT)
detector during the years 2014 and 2016. The HFT has demonstrated
excellent performance in reconstructing heavy flavor hadrons. We also
look forward to the measurement of $D^{0} (\bar{u}c)$ and
$\overline{D^{0}} (u\bar{c})$ directed flow utilizing both the HFT and ZDC
detectors in STAR to probe the early transient magnetic field.

\section{Acknowledgments}
This work is supported in part by the US Dept. of Energy under grant 
DE-FG02-89ER40531.

\end{document}